\documentclass[12pt]{article}  
\usepackage{amssymb,eufrak,latexsym}
\usepackage[mathscr]{eucal}

\newcommand{\g}{{\mathfrak g}}

\newcommand{\Z}{{\mathbb Z}}
\newcommand{\R}{{\mathbb R}}
\newcommand{\C}{{\mathbb C}}

\newcommand{\Lie}{{\mathcal L}}
\newcommand{\beq}{\begin{equation}}
\newcommand{\eeq}{\end{equation}}
\newcommand{\beqa}{\begin{eqnarray}}
\newcommand{\eeqa}{\end{eqnarray}}

\newcommand{\uno}{\mbox{1 \kern-.59em {\rm l}}}
\newcommand{\dalpha}{\dot{\alpha}}
\newcommand{\dbeta}{\dot{\beta}}

\newcommand{\nn}{\nonumber}
\begin{document}
\begin{titlepage}
\begin{flushright}
{ROM2F/2002/27}\\
{SISSA 75/2002/fm}\\
{PAR-LPTHE 02-54}\\
{STR-02-042}
\end{flushright}
\begin{center}
 
{\large \sc Multi-Instanton Calculus and Equivariant Cohomology}\\ 
  
\vspace{0.2cm}
{\sc Ugo Bruzzo}\\
{\sl Scuola Internazionale Superiore di Studi Avanzati, I.N.F.N. Sez. di Trieste\\ 
Via Beirut 4, 34013 Trieste, Italy} \\

{\sc Francesco Fucito}\\
{\sl Dipartimento di Fisica, Universit\'a di Roma ``Tor Vergata'',
I.N.F.N. Sezione di Roma II,\\
Via della Ricerca Scientifica, 00133 Roma, Italy}\\
{\sc Jose F. Morales}\\
{\sl Spinoza Institute,
Utrecht, The Netherlands\\
Laboratori Nazionali di Frascati, P.O. Box, 00044 Frascati, Italy}\\
and\\
{\sc Alessandro Tanzini}\\
{\sl Laboratoire de Physique Th\'eorique et Hautes Energies,\\
 Universit\'e Paris~VI, 4 Place Jussieu 75252 Paris Cedex 05, France}
\end{center}
\vskip 0.5cm
\begin{center}
{\large \bf Abstract}
\end{center}
{ We present a systematic derivation of multi-instanton amplitudes
in terms of ADHM equivariant cohomology. The results
rely on a supersymmetric formulation of the localization 
formula for equivariant forms. We examine the cases of ${\cal N}=4$
and ${\cal N}=2$ gauge theories with adjoint and fundamental matter.}

\par    \vfill
\end{titlepage}
\addtolength{\baselineskip}{0.3\baselineskip} 
\setcounter{section}{0}
\section{Introduction}                   
This year has seen dramatic improvements in our capabilities to handle multi 
instanton calculus. This is because powerful localization methods have been
applied to these computations.
 
The use of supersymmetric field theories to compute topological invariants 
was advocated in \cite{Witten:1988ze} and since then much work has been carried out
to clarify the formalism and its applications. In our opinion, all of this
work has not cleared a widespread belief that considered these methods not very relevant
for ``physical'' cases. This paper is about one of such cases: the computation
via path integral methods of non perturbative contributions due to instantons
for Yang-Mills gauge theories with ${\cal N}=2,2^*,4$ supersymmetries (SYM).

The computation of non perturbative effects has been the focus of much recent research.
Very often one studies such effects in the framework of string theory or supergravity and,
as a by product, recovers SYM with a low energy limit. Non perturbative
results for SYM thus provide important checks on these constructions 
like in the case of the AdS/CFT correspondence.

Notwithstanding its importance, the problem of extracting non perturbative results from
path integral computations has been not so intensively studied may be because every little
advance has been at the cost of a lot of effort. A short ``historical'' {\it excursus} will
clarify better what we mean.

To a ``primordial'' era in which the basic
techniques were estabilished \cite{Konishi:1989em}, aiming at supersymmetry breaking in gauge 
theories, it followed a period of stasis which was broken by the analysis of ${\cal N}=2$ SYM
carried out in \cite{sw}. To check this analysis, a lot of effort went in the 
computation of instanton effects
for winding numbers, $k$, larger than one \cite{DKM,ft}. Unfortunately these efforts were frustrated
by the fact that the ADHM constraints can be solved in an explicit way only for $k=1, 2$.
Explicit results could then be obtained only for the above mentioned values of $k$.
At this point it is worth to mention that in order to compare with the results obtained for the
prepotential in \cite{sw} there are two main directions: the first one employed in \cite{DKM}
relied on the computation of correlators which can be extracted directly from the effective 
Lagrangian and that exhibit a dependence on the space-time coordinates. 
The other, followed in \cite{ft}, was to compute the correlator $u=<Tr\varphi^2>$
where $\varphi$ is the complex scalar field of the ${\cal N}=2$ supersymmetric multiplet and $u$ is the
gauge invariant coordinate which describes the auxiliary Seiberg-Witten curve. Due to supersymmetric
Ward identities, $<Tr\varphi^2>$ is a pure number times a scale to the power of the naive dimension 
of the operator computed. The prepotential is then recovered using the identities of \cite{Matone:1995rx}.
The advantage of the second approach, in the light of the developments that have happened since then,
is that the computation of the correlator reduce to that of 
the partition function of a suitable matrix model defined on
the moduli space of instantons, whose action is generated by the presence 
of a vev for the scalar field.

Since those early efforts and the most recent developments, there has been two main advances:
on the one hand a measure for the moduli space of the instantons has been proposed which has led to many 
interesting results and that allows to write correlators without explicitly solving the constraints (for
a complete review see \cite{Dorey:2002ik}). On the other hand the agreement between the results of
\cite{DKM,ft}, which were obtained in the semiclassical expansion, and those of \cite{sw} which did not have
this shortcoming, triggered a subset of the present authors to recompute the correlators formulating the 
problem in the light of topological theories for which the semiclassical expansion is exact.
In \cite{Bellisai:2000bc} this program was carried out: the action, after the imposition of the constraint,
was written as the BRST variation of a suitable expression. The BRST charge thus defined squared to zero
after properly taking into account all the symmetries of the theory (see later for more details).
The first part of this program was completed in \cite{Bruzzo:2000di} in which the measure was shown to be 
BRST closed. The multi-instanton action ,meausure and the 
entire correlator of interest can then be written in a BRST exact form as it was shown in 
\cite{Dorey:2000zq} for ${\cal N}=4$ and in \cite{Fucito:2001ha} for ${\cal N}=2$. 

We now come to the last part of our story. Already in \cite{Bellisai:2000bc} the use of localization 
formulae was advocated in the study of multi instanton calculus. But at that time it was not clear to 
the authors how to deal with the singularities of the instanton moduli space and with its boundaries.
In \cite{Bellisai:2000bc}, in fact, a formula was given for the correlators as boundary terms over the 
moduli spaces of $SU(2)$ instantons but explicit computations could be carried out only in the $k=1$ case.

In \cite{Dorey:2000zq} it appeared the proposal to deform the moduli space by minimally resolving the 
singularity  using the invariance, under certain 
deformations, of the original BRST exact theory.
Moreover, in the same paper, it was also suggested that the action had its minima in certain points that 
could be interpreted as resolved Hilbert schemes\cite{nakajima}\footnote{Sometimes, in some physics 
literature, they are
called improperly ``$U(1)$ instantons''. We prefer the definition appearing in the mathematical literature.}.
These ideas were then coherently applied in \cite{Hollowood:2002ds} in which the author 
recomputed the $k=1, 2$ cases showing the full applicability of localization techniques to this problem.

There was a last crucial ingredient which was missing and it was provided in \cite{Nekrasov:2002qd}: 
localization techniques
are most powerful when the critical points of the action are isolated. In order to have such
isolated points the
BRST transformations of the theory must be further deformed by  a rotation in the space-time which,
in turn, induces an action on the moduli space. This action leaves the ADHM constraints invariant. The
cohomology of the BRST operator is thus the same of that of the undeformed theory and it can be used safely
if we find it to be more convenient. This technique has been used widely in the mathematical literature.
It has also appeared in the physics literature, his most notable applications being the computation of
the D-instanton partion function\cite{Moore:1998et} and the study of the role of momentum 
maps for supersymmetric theories\cite{Moore:1997dj,Losev:1997tp}. 
This stream of physics literature might be traced 
back to the
investigations of the possible use of the Duistermaat-Heckman formula in the context
of supersymmetric gauge theories in the two dimensional case\cite{Witten:1992xu}. This will be a 
particular case of the most general situation we are going to treat in the next section.

At last we come to the content of this paper: we share with \cite{Flume:2001kb,Flume:2002az}, 
the belief that the
ideal setting for these computations is that of equivariant cohomology (see\cite{berline} for a complete 
review). In the ${\cal N}=2$ case the formulae found in \cite{berline} give the correct result because in this case 
the number of fermions and bosons are the same, and the fermions belongs to the tangent space of
the bosonic moduli.
There is also a way to avoid treating the constraints, by 
modifying some computations in \cite{nakajima}; but in the general case these formulae have to be modified.
This is what we do here, by discussing the results that can be obtained from a proper supersymmetric 
formulation of the localization formulae for equivariant forms along the lines of \cite{Bruzzo:2000di}. 
Full details on this extension will appear elsewhere \cite{bf}.

This is the plan of the paper:
in the next section we carefully define and explain the objects which will enter the localization 
formula in the case of the bosonic theory. We keep the discussion as elementary as possible giving
examples each time we introduce new objects. 
In the third and last section we first identify the objects introduced in the previous section as the 
building blocks of supersymmetric gauge theories.
We then compute the  ${\cal N}=2, 2^*, 4$ cases. For the sake of clarity we have decided to 
relegate mathematical considerations and comparisons with previous literature to the appendices.

\section{Preliminaries \label{intro}}
\setcounter{equation}{0} 
\subsection{The ADHM data}

The moduli space of self-dual solutions of $U(N)$ YM-equations 
in four dimensions is elegantly described by a $4kN$-dimensional 
hypersurface embedded in a $4kN+4k^2$-dimensional ambient space
via ADHM constraints.
In the case of supersymmetric gauge theories the ADHM data are
supplemented with fermionic moduli associated to zero modes of
the gaugino field. The multi-instanton action is 
defined by plugging in the SYM lagrangian the bosonic and fermionic
zero modes in terms of ADHM moduli and imposing the ADHM constraints
via lagrangian multipliers. 

There is a quick way to perform all of these steps at once and it is to use the corresponding
D-brane description. 
The ADHM action for ${\cal N}=4$ can be extracted from the low energy dynamics of a system
of $k$ D(-1) and $N$ D3-branes moving in flat space
\cite{Douglas:1995bn,Douglas:1996uz,Witten:1995gx,Akhmedov:1998pf}, 
the moduli of the four dimensional 
supersymmetric theory being the massless excitations of the open strings stretching between various branes.
The complete ADHM Lagrangian can, in fact, also be derived from the computation of disks amplitudes 
in string theory \cite{Green:2000ke,bfpfll}.  

This results into a zero-dimensional quantum theory of matrices,
some transforming in the adjoint of $U(k)$ and others in the bifundamental
of $U(N)\times U(k)$. 
Less supersymmetric multi-instanton actions are then found via suitable
projections of the original ${\cal N}=4$ theory (see below for details). 

Let us  start by describing the ${\cal N}=4$ ADHM data \cite{Dorey:2000zq,dhkmv}. 
The position 
of $k$ D(-1)-instantons in ten-dimensional space can be described by
five complex fields $B_{\ell},\phi$ with $\ell=1,..4$. For latter
convenience we have distinguished one of the complex 
planes and denoted it by $\phi$. In addition open strings stretching
between D(-1)-D3 branes provide two extra complex moduli $I,J$ in the
$(\bar{k},N)$ and $(\bar{N},k)$ bifundamental representations respectively
of $U(k)\times U(N)$. The $U(N)$ group, together with the $SO(4)\times SO(6)$ 
Lorentz group preserved by the D(-1)-D3 system,
act as the group of global isometries of the $U(k)$ zero-dimensional
quantum theory living in the D(-1)-worldvolume. 
Supersymmetry requires bosonic moduli to be paired with  
fermionic ones. Bosonic ADHM moduli in the 
adjoint of $U(k)$ come together with sixteen fermionic components 
$(\chi_v,\eta,{\cal M}_{\ell},\bar{\cal M}_{\ell})$, $v=1,\ldots ,7$, $\ell=1\ldots , 4$,
coming from the reduction under a subgroup $SO(2)\times SO(7)$ \cite{Hirano:1997ai}
of a single Majorana-Weyl
fermion in $D=10$ down to zero dimensions. 
 Again the splitting of the fermionic
fields in $7+1$ real and $4$ complex components is a matter of convenience.
 
 On the other hand fermionic excitations of D(-1)-D3 open strings
provide two pairs $(\mu_I,\mu_K)$,  $(\mu_J,\mu_L)$ of complex
fields in the $(\bar{k},N)$ and $(\bar{N},k)$ bifundamental representations
respectively.

 The last ingredient in the construction of the instanton moduli
space is the ADHM constraints. 
ADHM constraints can be efficiently implemented by Lagrangian multipliers. 
For this purpose it is convenient to introduce the auxiliary fields
$K,L,H_{\R},H_r$ with $r=1,2,3$. 
To the adjoint auxiliary fields $H_{\R},H_r$ we associate respectively
one real and three complex functions 
\beqa\label{mommap4}
{\cal E}^{\rm adj}_{\R}&=&[B_\ell,B_\ell^\dagger]+II^\dagger-J^\dagger J -\zeta
 \ \ , \nonumber\\
{\cal E}^{\rm adj}_{1}&=&[B_1,B_2]+[B_3^\dagger, B_4^\dagger]+IJ\ \ ,\nn\\
{\cal E}^{\rm adj}_{2}&=&[B_1,B_3]-[B_2^\dagger, B_4^\dagger] \ \ ,\nn\\
{\cal E}^{\rm adj}_{3}&=&[B_1,B_4]+[B_2^\dagger, B_3^\dagger]  \ \ ,
\eeqa
and to the fundamental auxiliary fields $(K,L)$ the two complex functions
\beqa\label{mommap5}
{\cal E}^{\rm fun}_{K}&=&B_3 I-B_4^\dagger J^\dagger \nn\\
{\cal E}^{\rm fun}_{L^\dagger}&=&B_4 I+B_3^\dagger J^\dagger  \ \ .
\eeqa
As we will see in the next section, after reduction to ${\cal N}=2$,
(\ref{mommap4}) will reduce to the familiar ADHM constraints.
For greater generality we have added a non commutativity parameter 
$\zeta$ to minimally resolve the small instantons
singularities of the moduli space.
In the following we will collect the $3$ complex and $1$ real
adjoint components in (\ref{mommap4})
in the seven-vector ${\cal E}^{\rm adj}_v$ , $v=1,\ldots,7$,  
and denote the associated auxiliary fields as $\chi_v,H_v$.

 The ADHM data just described
can be nicely organized in multiplets of a BRST current $Q$ \cite{Dorey:2000zq}
\footnote{See \cite{Bellisai:2000bc,Fucito:2001ha} for the ${\cal N}=2$ case}. 
$Q$ is a BRST current in the sense that it squares to zero up to a 
$U(k)$ gauge transformation on the moduli space. We will need
a slight modification of this BRST charge, in which 
the $U(k)$ group action is combined with 
an element of a $U(1)^{N-1}\times U(1)^3$ subgroup in the 
$SU(N)\times SO(4)\times SO(6)$ global symmetry group of the
D(-1)-D3 system. This choice is dictated by the requirement that the
BRST charge closes up to a group action with isolated critical points.
As firstly appreciated in  \cite{Nekrasov:2002qd} this allows to reduce
integrals in the ADHM moduli space to a sum over critical points.   
In the next subsection we will state our main localization formula
for group actions meeting this basic requirement.  
The reader interested in a deeper understanding of the discussion
in this section and the connection with previous results in 
\cite{Bellisai:2000bc,dhkmv} is referred to the Appendix B. 

We denote the new BRST charge by $Q_\epsilon$ and parametrize an
element in ${\cal T}=U(1)^{N-1}\times U(1)^3$ by 
$a_\lambda,\epsilon_1,\epsilon_2,m$ with $\lambda=1,\ldots,N$ and $\sum a_\lambda=0$.
The $U(1)_{\epsilon_{1,2}}$'s are inside the $SO(4)$ Lorentz
group while $U(1)_m$ is chosen in $SO(6)$. The m-deformation breaks
the $SO(6)\sim SU(4)$ ${\cal R}$-symmetry group of ${\cal N}=4$ down to
the $SU(2)\times U(1)$ ${\cal R}$-symmetry group of ${\cal N}=2$. 
The ${\cal N}=4$ adjoint vector multiplet decomposes into a 
vector and an hypermultiplet of ${\cal N}=2$.   
Later we will identify the parameter $m$ with the mass of the 
hypermultiplet. Keeping this identification in mind we call
the deformed ${\cal N}=4$ theory as ${\cal N}=2^*$.
Pure ${\cal N}=2$ SYM theory can instead be defined
by a $\Z_2$-projection with $\Z_2\subset U(1)_m$.

Given all of this, the deformed ${\cal N}=2^*$ multi-instanton action 
can be
written as the $Q_\epsilon$-exact form \cite{Dorey:2000zq}:
\beq
S^{{\cal N}=2^*}= Q_{\epsilon}\, {\rm Tr} \left[{1\over 4}\eta[\phi,\bar\phi]+
\vec H\cdot\vec\chi-i\vec{\cal E}\cdot \vec \chi-{1\over 2}
\sum_{s=1}^6(\Psi_s^\dagger(\bar\phi+\lambda_s)\cdot B_s+
\Psi_s (\bar\phi+\lambda_s)\cdot B_s^\dagger)\right]
\label{N=2*act}
\eeq 
with $B_s =(I,J^\dagger,B_\ell)$,
$\Psi_s=(\mu_I,\mu_J^\dagger,{\cal M}_\ell)$, 
$\vec{\chi}=(\mu_K,\mu_{L^\dagger},\chi_v)$, 
$\vec{H}=(K,L^\dagger,H_v)$ and
$\vec{{\cal E}}=({\cal E}^{\rm fun}_K,
{\cal E}^{\rm fun}_{L^\dagger},{\cal E}^{\rm adj}_v)$.
The convention for the vector product is 
$\vec{\chi}\cdot\vec{H}\equiv {1\over 2} \chi_{\R}H_{\R} + \chi_{r}^{\dagger}H_{r}
+ \chi_{r}H_{r}^{\dagger}$, while $\phi\cdot B_s=[\phi,B_s]$ or 
 $\phi\cdot B_s=\phi B_s$ depending or whether $B_s$ is in the 
$U(k)$ adjoint or fundamental representation respectively .

The BRST transformations are given by:
\beqa\label{brsn4}
Q_\epsilon I&=&\mu_I\ \;\qquad Q_\epsilon\mu_I=\phi I-I a\nonumber\\
Q_\epsilon J&=&\mu_J\ \;\qquad Q_\epsilon\mu_J=-J\phi+a J 
+\epsilon J  \nonumber\\
Q_\epsilon \mu_K&=&K\ \;\qquad Q_\epsilon K=\phi \mu_K-\mu_K a -m \mu_K\nonumber\\
Q_\epsilon \mu_L&=&L\ \;\qquad Q_\epsilon L=-\mu_L\phi +a \mu_L +(\epsilon-m) \mu_L 
 ,\nonumber\\
Q_\epsilon B_\ell&=&{\cal M}_\ell \ \;\qquad 
Q_\epsilon{\cal M}_\ell=[\phi,B_\ell]+\lambda_\ell B_\ell\  \nonumber\\
Q_\epsilon \chi_v&=&H_v\ \;\qquad Q_\epsilon H_v=[\phi,\chi_v]+\lambda_v \chi_v\ \nn\\
Q_\epsilon\bar\phi&=&\eta \ \;\qquad 
Q_\epsilon\eta=[\phi,\bar\phi] \,\nonumber\\
Q_\epsilon\phi&=&0\ \ ,
\eeqa
with $\epsilon=\epsilon_1+\epsilon_2$, $\phi\in U(k)$ and
\beqa\label{lambdas}
a &=&{\rm diag}\,(e^{i a_1},e^{i a_2}, \ldots e^{i a_N})\nn\\
\lambda_s&=&(\lambda_I,\lambda_{J^\dagger};\lambda_\ell) 
=(0,-\epsilon;\epsilon_1,\epsilon_2,-m,m-\epsilon)\nn\\
\lambda_v &=&(\lambda_{\R};\lambda_r)=
(0;\epsilon,\epsilon_1-m,m-\epsilon_2)
\eeqa
The group assigments in the
right-hand side of (\ref{brsn4}) are determined by the requirement
that auxiliary fields associated to
 (\ref{mommap4}) transform covariantly under  
the $U(k)\times SU(N)\times U(1)^3$ transformations:
\beqa\label{invmod4}
I&\rightarrow& g_{U(k)}\, I\, g^\dagger_{U(N)} \nonumber\\
J&\rightarrow& e^{i\epsilon}\, g_{U(N)}\, J\, g^\dagger_{U(k)} \nonumber\\
K &\rightarrow& e^{-i m}\, g_{U(k)}\, K\, g^\dagger_{U(N)} \nonumber\\
L &\rightarrow& e^{i(m-\epsilon)}\, g_{U(N)}\, L\, g^\dagger_{U(k)} \nonumber\\
B_\ell&\rightarrow& e^{i\lambda_\ell}\, g_{U(k)}\,B_\ell\, g^\dagger_{U(k)}
 \nn\\
H_v&\rightarrow& e^{i\lambda_v}\, g_{U(k)}\, H_v\, g^\dagger_{U(k)} 
\eeqa
with $\lambda_\ell,\lambda_v$ given by (\ref{lambdas}) and similar 
expressions for the fermionic superpartners.

\subsection{Equivariant Forms and the Localization Formula}

Let $M$ be an $n$-dimensional manifold acted on by a Lie group $G$ with Lie 
algebra $\g$.  For every $\xi\in\g$ we denote by
$\xi^*$  the \emph{fundamental vector field} associated
with $\xi$, i.e., the vector field that generates the one-parameter group 
$e^{t \xi}$ of transformations of $M$.
Locally one has 
$$\xi^* =\xi^\alpha \,T_{\alpha}^i\frac{\partial}{\partial x^i}$$ 
where the $\xi^\alpha$ are the components of $\xi$ in some chosen basis
of $\g$, and the $T_\alpha^i$ are functions (the generators of the action).

Let $\alpha:\g\rightarrow \Omega(M)$ a polynomial map from $\g$ to the algebra
of differential forms on $M$. $\alpha$ may be regarded as an element of
$\C[\g]\otimes\Omega(M)$ with $\C[\g]$  the algebra 
of complex-valued polynomials on $\g$.
We define a grading in $\C[\g]\otimes\Omega(M)$ by letting, for homogeneous
$P\in\C[\g]$ and $\beta\in\Omega(M)$, 
\beq
\deg(P\otimes\beta)=2 \deg(P)+\deg(\beta).
\label{deg}
\eeq
The action of the group $G$ on an element 
$\alpha\in\C[\g]\otimes\Omega(M)$ is defined to be 
\beq
(g\cdot\alpha)(\xi)=g^\ast(\alpha(Ad_{g^{-1}}\xi))
\label{gaction}\eeq 
where $g^\ast$ denotes the pullback of forms with respect to
the map $g\colon M\to M$.
Elements $\alpha$ such that $g\cdot\alpha=\alpha$ are called
$G$-equivariant forms. 
The  equivariant differential ${\cal D}$ is defined  by letting 
\beqa
({\cal D}\alpha) (\xi) &=&d(\alpha(\xi))-i_{\xi^*}\alpha(\xi)
\label{der}\eeqa
where $i_{\xi^*}$ is the inner product by the vector field $\xi^*$.

As an example let us take $G=O(2)$,
$M=\R^2-\{0\}$ with the standard action of $G$, and $$\alpha(\xi)=\xi r d\theta$$
(we write the matrices in the Lie algebra $\mathfrak{so}(2)$ as $0\ \ \xi \choose-\xi \ 0$).
Explicit computation shows that $\alpha$ is equivariant. Since  
\beq \xi^*= 2\xi \frac{\partial}{\partial \theta}\eeq
 one has
\beq
({\cal D}\alpha) (\xi)=d(\alpha(\xi))-i_{\xi^*}\alpha(\xi)=-\xi
 dr\wedge d\theta-\xi r^2.
\label{derex}\eeq
In the first term on the r.h.s. of (\ref{derex}) the degree of the one-form $\alpha(\xi)$ has been raised
by one unit, while in the second term    it has been lowered by one unit. But in this 
very term the degree of the polynomial in $\C[\g]$ has been raised by one 
and therefore according to (\ref{deg})
the total degree is raised by one unit, as we expect from a 
derivation. 

Acting twice on a $G$-equivariant form $\alpha$ one finds:
 \beq
({\cal D}^2 \alpha)(\xi)=(d-i_{\xi^*})^2 \alpha(\xi)=
-(di_{\xi^*}+i_{\xi^*}d)\alpha(\xi)=- \Lie_{\xi^*}\alpha(\xi)=0
\label{der2}\eeq
since $\alpha$ is equivariant.
The space of equivariant differential forms with this differential, graded
with the degree (\ref{deg}), is
a differential complex; its cohomology is called the $G$-equivariant cohomology of $M$.

We shall denoted by $\alpha_i(\xi)$ the
homogeneous component of degree $i$ of the differential
form $\alpha(\xi)$. The condition that $\alpha$ is equivariantly
closed, ${\cal D}\alpha=0$, implies that $\alpha_n(\xi)$ (with $n=\dim M$) is exact 
outside of the set $M_0$ of zeros of $\xi^*$ \cite{berline}, suggesting that the integral
$\int_M \, \alpha(\xi)$ reduces to an integral over 
$M_0$. This is the content of the localization formula below.

Let now $x_0$  be a  zero of $\xi^*$. We introduce a map 
${\cal L}_{x_0}:T_{x_0} M\to T_{x_0} M$  defined as 
\beq\label{unoapp}
{\cal L}_{x_0}(v) =[\xi^*,v]= 
-\xi^\alpha\, v^i\left(\frac{\partial T_\alpha^j}{\partial x^i}\right)_{x_0}\frac{\partial}{\partial
x^j},
\eeq
(which makes sense because at the critical points the components of the fundamental vector field vanish,
$\xi^\alpha T_\alpha^i(x_0)=0$).

In particular cases ${\cal L}_{x_0}$ can be 
interpreted as a Hessian; this is the case of Morse theory. But a Hessian is defined given a certain
reference function to be derived twice. On the contrary ${\cal L}_{x_0}$  is known once we know  
the group  action.  The reader should keep this in mind since this  is a fact which will be of great
relevance  in the next section. 

Given all of this, assuming that both $M$, $G$ are compact, 
$\alpha$ equivariantly closed, and that $\xi\in\g$ is such that
the vector field $\xi^*$  has only isolated zeroes, 
we can state the localization 
theorem\footnote{The interested
reader can consult 
\cite{berline} for a proof.} 
\beq
\int_M \alpha(\xi)=(-2\pi)^{n/2}\sum_{x_0}{\alpha_0(\xi)(x_0)\over 
{\rm det}^{1\over 2}\, {\cal L}_{x_0}}.
\label{locth}
\eeq
By $\alpha_0(\xi)$ we mean the part of $\alpha(\xi)$ which is a zero-form.  

There is a nice supersymmetric formulation of this equation. 
A complete proof of this result
will be presented in \cite{bf}, here we just state the result.
Let $\mathfrak M$ be a $(n,n)$-dimensional supermanifold, defined
in such a way that the superfunctions on $\mathfrak M$ are
(non-homogeneous) differential forms. This condition in particular
implies a splitting $T_{x_0}\mathfrak M = T_{x_0}M\oplus T_{x_0}M$,
and one can introduce a (odd) linear transformation $\Pi\colon 
T_{x_0}\mathfrak M\to T_{x_0}\mathfrak M$ defined by exchanging the
two copies of $T_{x_0}M$. If   $(x^i)$ is a local coordinate
chart on $M$, and $(\theta^j)$ a local basis of differential 1-forms, then
$(x^i,\theta^j)$ is a local coordinate chart on the supermanifold
$\mathfrak M$. 

The group $G$ acts naturally on $\mathfrak M$, extending the action on $M$.
For every $\xi\in\g$  one has an even supervector field on $\mathfrak M$,
\beq\hat\xi^* = \xi^\alpha\,T_\alpha^i\,{\partial\over \partial x^i} 
+ \xi^\alpha\,\theta^j \,\frac{\partial T_\alpha^i
}{\partial x^j}\,\frac{\partial}{\partial \theta^i}.\label{recipe}\eeq
One can  also
introduce an odd vector field $Q^*$ on $\mathfrak M$, defined as
\beq Q^* =
\theta^i\,\frac{\partial}{\partial x^i} + 
  \xi^\alpha \,T_\alpha^i \,\frac{\partial}{\partial \theta^i} \,.\label{BRST} \eeq
 A simple computation
shows that \emph{the anticommutator of $Q^*$ with itself is twice the generator $\hat\xi^*$}:
$${1\over 2}\{ Q^*,Q^* \}=\hat\xi^*.$$
The vector field $Q^*$ here may be regarded as the \emph{infinitesimal generator
of the BRST transformations}.

The localization formula can now be stated as follows
\beq
\int_{M} \alpha(\xi)=(-2\pi)^{n\over 2}
\sum_{x_0} {\alpha_0(\xi)(x_0)\over \vert {\rm Sdet}'\, {\cal L}_{x_0}\vert^{1\over 2}} .
\label{locths}\eeq
where now 
${\cal L}_{x_0}: 
T_{x_0}\mathfrak M \to T_{x_0}\mathfrak M$ is given by ${\cal L}_{x_0}=[Q^*,v]$,
and the operator $\mbox{Sdet}'$ is defined as $\mbox{Sdet}\circ\Pi$. With this provision,
from now on each time we write the superdeterminant we intend it to be primed.

\section{Applications to Supersymmetric Gauge Theories \label{conti}}
\setcounter{equation}{0} 

From (\ref{N=2*act}) we conclude that the ${\cal N}=4$ multi-instanton
action and their descendants upon mass deformation, ${\cal N}=2^*$,
or orbifold projection, ${\cal N}=2$, are equivariantly exact
forms. Together with the $U(k)\times U(N)\times U(1)^3$-invariance this
implies that they are equivariantly closed.
We can thus apply the localization techniques explained in the previous section
to compute the multi-instanton
partition function for all these theories.
The Appendix B collect some background material that can help for a deeper understanding of the
results presented in this section.  

\subsection{${\cal N}=2$ Supersymmetric Theories with Gauge Group $SU(N)$}

We start discussing the
case of ${\cal N}=2$ gauge theories with $N_F$ fundamental hypermultiplets.
Let us first recall the content of the ${\cal N}=2$ 
ADHM data in the pure ${\cal N}=2$ case $N_F=0$.
 Pure ${\cal N}=2$ SYM theory can be obtained
by placing a stack of $N$ fractional D3-branes at a $\R^4/\Z_2$ singularity.
The ADHM instanton moduli space can then be described by implementing
the $\Z_2$-projection on the D(-1)-D3 system describing the ${\cal N}=4$ parent
theory  \cite{Fucito:2001ha}.
The net effect of such operation is to break the $SU(4)$ 
${\cal R}$-symmetry group of the ${\cal N}=4$ theory down to 
$SU(2)_{\dot{A}}\times U(1)_{\cal R}$,
where $SU(2)_{\dot A}$ is the ${\cal N}=2$ automorphism 
group and $U(1)_{\cal R}$ the anomalous ${\cal R}$ charge. 
Choosing a $\Z_2\subset U(1)_m$, the projection
corresponds to set to zero all the fields 
charged under $U(1)_m$ in 
(\ref{invmod4})
\beq
B_3=B_4=K=L=H_2=H_3=0
\label{riduz}
\eeq 
together with their fermionic superpartners.
The field content is thus reduced to 
$B_{\hat{s}} =(I,J^\dagger,B_{\hat{\ell}})$,
$\Psi_{\hat{s}}=(\mu_I,\mu_J^\dagger,{\cal M}_{\hat{\ell}})$,
with $\hat{s}=1,\ldots,4$,  $\hat{\ell}=1,2$, and 
$\vec{\chi}=\chi_{\hat{v}}$ and $\vec{H}=H_{\hat{v}}$
with $\hat{v}=1,2,3$.

The ${\cal N}=2$ multi-instanton action is then obtained from (\ref{N=2*act})
by simply replacing $s,v$ with $\hat{s},\hat{v}$
\beq
S^{{\cal N}=2}= Q_{\epsilon}\, {\rm Tr} \left[{1\over 4}\eta[\phi,\bar\phi]+
\vec H\cdot\vec\chi-i\vec{\cal E}\cdot \vec \chi-{1\over 2}
\sum_{\hat{s}=1}^4(\Psi_{\hat{s}}^\dagger(\bar\phi+\lambda_{\hat{s}})\cdot 
B_{\hat{s}}+
\Psi_{\hat{s}} (\bar\phi+\lambda_{\hat{s}})\cdot B_{\hat{s}}^\dagger)\right]
\label{actionn=21}
\eeq 
In particular the surviving equations in (\ref{mommap4})
reproduce the familiar ADHM constraints:
\beqa\label{mommap2}
{\cal E}_\R&=&[B_1,B_1^\dagger]+[B_2,B_2^\dagger]+II^\dagger-J^\dagger J-\zeta=0
\nonumber\\
{\cal E}_\C&=&[B_1,B_2]+IJ=0 
\eeqa
The presence in (\ref{mommap2}) of the
non commutativity parameter $\zeta$ allows to minimally resolve the orbifold
singularities of the moduli space.
The action (\ref{actionn=21}) 
is invariant under
\beqa\label{brsn2}
Q_\epsilon I&=&\mu_I\ \;\qquad Q_\epsilon\mu_I=\phi I-I a\nonumber\\
Q_\epsilon J&=&\mu_J\ \;\qquad Q_\epsilon\mu_J=-J\phi+a J 
+\epsilon J  \nonumber\\
Q_\epsilon B_{\hat{\ell}}&=&{\cal M}_{\hat{\ell}} \ \;\qquad 
Q_\epsilon{\cal M}_{\hat{\ell}}=[\phi,B_{\hat{\ell}}]+\epsilon_{\hat{\ell}} 
B_{\hat{\ell}}\  \nonumber\\
Q_\epsilon \chi_{\hat v}&=&H_{\hat v}\ \;\qquad 
Q_\epsilon H_{\hat v}=[\phi,\chi_{\hat v}]+\lambda_{\hat v} \chi_{\hat v} \nn\\
Q_\epsilon\bar\phi&=&\eta \ \;\qquad 
Q_\epsilon\eta=[\phi,\bar\phi] \,\nonumber\\
Q_\epsilon\phi&=&0\ \ .
\eeqa
with $\hat{\ell}=1,2$ and $\lambda_{\hat v}=(\lambda_\R,\lambda_\C)=(0,\epsilon)$.

We would like now to apply the localization formula (\ref{locths}) to
compute the partition function in the 
multi-instanton moduli space described by the 
bosonic  ${\cal B}=(I,J,B_{\hat{\ell}},H_{\hat{v}},\bar\phi)$ and 
fermionic ${\cal F}=(\mu_I,\mu_J,{\cal M}_{\hat{\ell}},
\chi_{\hat{v}},\eta)$ variables.

To this end we start by introducing 
the vector field
$Q^*$ generating the BRST transformations on the supermanifold
and discussing the critical points of its action.
From (\ref{brsn2}) we get
\beqa\label{fermvecf}
Q^* &=&\mu_I{\partial\over\partial I}+
\mu_J{\partial\over\partial J}+
{\cal M}_{\hat{\ell}}{\partial\over\partial B_{\hat{\ell}}}
+H_{\hat{v}}{\partial\over\partial \chi_{\hat{v}}}
+\eta{\partial\over\partial \bar\phi}
+([\phi,\chi_{\hat{v}}]+\lambda_{\hat v}\chi_{\hat{v}}){\partial\over\partial H_{\hat{v}}}+
[\phi,\bar\phi]{\partial\over\partial \eta}\nonumber \\
&&(\phi-a)I{\partial\over\partial \mu_I}+(-\phi +a+\epsilon)J
{\partial\over\partial \mu_J}+
([\phi,B_{\hat{\ell}}]+\epsilon_{\hat{\ell}} B_{\hat{\ell}})
{\partial\over\partial {\cal M}_{\hat{\ell}}}
\nonumber \\
&=&
(Q^*)^i_{\cal B}{\partial\over\partial {\cal B}^i}+(Q^*)^i_{\cal F}{\partial\over\partial {\cal F}^i},
\eeqa
The critical points $Q^*=0$ are given by setting to zero the components
in (\ref{fermvecf})
\beqa \label{critical}
&&(\varphi_{IJ}+\epsilon_\ell)\, B^\ell_{IJ}=0 \nn\\
&& (\varphi_{I}-a_\lambda)\, I_{I\lambda}=0 \nn\\
&& (-\varphi_{I}+a_\lambda+\epsilon)\, J_{\lambda I}=0
\eeqa
with $H_{\hat{v}}$ and all fermions set to zero. $H_{\hat{v}}=0$
implements the ADHM constraints (\ref{mommap2}).
Eqs. (\ref{critical}) were solved in \cite{Nekrasov:2002qd}. Each critical point
is associated to a set of $N$ Young Tableaux $(Y_1,\ldots Y_N)$ with 
$k=\sum_\lambda k_\lambda$ boxes distributed between the $Y_\lambda$'s.
The boxes in a $Y_\lambda$ diagram are
labeled either by the 
instanton index $I_\lambda=1,\ldots,k_\lambda$ or by the pair of integers $i_\lambda,j_\lambda$ 
denoting the vertical and horizontal
position respectively in the Young diagram.
We denote by $\nu_{i_\lambda},\nu^\prime_{j_\lambda}$ the
length of the $i_\lambda$-th row and $j_\lambda$-th column respectively. 
The solution can then be written as:      
\beqa \label{solcritical}
\varphi_{I_\lambda}=\varphi_{i_\lambda j_\lambda} &=& 
a_{\lambda}-(j_\lambda-1)\epsilon_1-
(i_\lambda-1)\epsilon_2 
\eeqa
and $J=B_\ell=I=0$ except
for the components $B^1_{(i_\lambda j_\lambda+1),(i_\lambda j_\lambda)}$,  
$B^2_{(i_\lambda+1 j_\lambda),(i_\lambda j_\lambda)},I_{\lambda,(i_\lambda=j_\lambda=1)}$. 
These apparent moduli correspond to the zero eigenvalues in the
left hand side of (\ref{critical}). They are however eliminated by the
ADHM constraints (\ref{mommap2}). This can be seen as follows:
at the critical points
\beq\label{ancrit}
{\cal E}_{\C}=[B_1,B_2]=0
\eeq
is non-trivial only when the box labelled by the pair $(i_\lambda j_\lambda)$
has a neighbor both on its left and its down direction.
These equations can then be used to determine, for example the corresponding components of $B_2$.
This leaves $\nu_{1_\lambda}-1$ undetermined components for $B_2$. In addition we have 
$k_\lambda-\nu_{1_\lambda}$
non-trivial $B_1$ components and one component for $I$. All together this leaves $k$ components
which are fixed by the diagonal components of the real constraint 
\beq\label{realconstrainfixed}
{\cal E}_\R=[B_1,B_1^\dagger]+[B_2,B_2^\dagger]+II^\dagger-\zeta=0.
\eeq
We conclude that critical points of the $U(1)^k\times U(1)^{N-1}\times
U(1)^2$
action are isolated.
We are now ready to apply the localization formula.
As in
\cite{Nekrasov:2002qd} we can use the $U(k)$-invariance to write
the $Q_\epsilon$-unpaired field $\phi$ as
$\phi_{IJ}=\varphi_I-\varphi_J$ in terms of $k$ $\varphi_I$ 
phases.
The Jacobian of this change of variables brings the so called
Vandermonde determinant $\prod_{I<J} \varphi^2_{IJ}$.
 According to our localization formula (\ref{locths}), we find
\beq\label{theformula}
{\cal Z}_k=\int {{\cal D}\phi\over U(k)} 
{\cal D}{\cal{B}} {\cal D}{\cal{F}} e^{-S}=
\int \prod_{I=1}^k d\varphi_I{\prod_{I<J} \varphi^2_{IJ}\over {\rm Sdet}\,{\cal L}}
\equiv \sum_{x_0}  {1\over {\rm Sdet}
 \hat{{\cal L}}_{x_0}}  \ \ ,
\eeq
having used the fact that the action $S^{{\cal N}=2}$ vanishes on the critical points,
thus $\alpha_0(\xi)(x_0)=1$.
The superdeterminant\footnote{Since our ADHM variables are complex, 
our tangent space is complex too and the determinant
to the inverse of the square root becomes simply the inverse of the determinant in the 
localization formulae.} is defined by 
\beq\label{sdet}
{\rm Sdet} {\cal L}=
Sdet\pmatrix{{\partial(Q^*)^i_{\cal B}\over\partial{\cal F}^j} & 
{\partial(Q^*)^i_{\cal B}\over\partial{\cal B}^j}\cr
{\partial(Q^*)^i_{\cal F}\over\partial{\cal F}^j} &
{\partial(Q^*)^i_{\cal F}\over\partial{\cal B}^j}}
\eeq
Plugging (\ref{sdet}) in (\ref{theformula}) one recovers (3.10) in \cite{Nekrasov:2002qd}
where the integral is computed in the complex plane with poles at the critical points (\ref{solcritical}).
The explicit form of the residue formula obtained in this way is however
difficult to handle. In the following we shall adopt the approach of \cite{Flume:2001kb}\footnote{
We thank R.Flume and R.Poghossian for detailed explanations of their work.}
that generalize to $U(N)$ the analysis
performed by Nakajima \cite{nakajima} 
in the study of resolved Hilbert schemes.
Following \cite{Flume:2001kb}  we start by computing the character 
$\chi\equiv \sum_i (-)^F\, e^{i \lambda_i}$, where the $\lambda_i$'s are
the eigenvalues of ${\cal L}_{x_0}$ and $(-)^F=\pm 1$ according to the gradation 
given by (\ref{fermvecf}).

As we will see the resulting character $\chi$ can be reduced  
via algebraic manipulations to a sum over $2kN$ eigenvalues.
The determinant is then found by replacing the sum by a product
over the $2kN$ eigenvalues. 

Notice that the extension of the localization formula to the superspace
allows to easily handle the linearized ADHM constraints by introducing the
fermionic ``ghost'' variables $(\chi_{\R},\chi_{\C})$.
As we will shortly see, in the computation one can in fact nicely recognize the
cancellations between bosonic and fermionic contributions that mimics
the reduction via ADHM contraints to the $2kN$-dimensional moduli space
(see Appendix C for details). This is a general feature of the superspace approach,
not necessarily linked to space-time supersymmetry.

Let us introduce the generators $T_\ell=e^{i\epsilon_\ell},
T_{a_\lambda}=e^{i a_\lambda}$ for elements in
$U(1)^{N-1}\times U(1)^2$. In addition we write
$V=e^{i \varphi_I}$ with $\varphi_I$ given by  (\ref{critical}).
The Supertrace of $\hat{{\cal L}}_{x_0}$ at the critical 
point (\ref{critical}) can be written (see Appendix C for a more detailed 
explanation) 
as
\beq \label{trace}
\chi=V^*\times V\times \left[T_1+T_2-T_1T_2-1\right]
+ W^*\times V
+V^*\times W\times T_1T_2
\eeq 
with 
\beqa\label{Vs}
V&=&\sum_{\lambda =1}^N\sum_{j_\lambda=1}^{\nu_{1_\lambda}}
\sum_{i_\lambda=1}^{\nu^\prime_{j_\lambda}}T_1^{-j_\lambda+1}
T_2^{-i_\lambda+1}T_{a_\lambda}=
\sum_{\lambda =1}^N\sum_{i_\lambda=1}^{\nu^\prime_{1_\lambda}}
\sum_{j_\lambda=1}^{\nu_{i_\lambda}}T_1^{-j_\lambda+1}T_2^{-i_\lambda+1}T_{a_\lambda},\nonumber\\
W&=&\sum_{\lambda =1}^nT_{a_\lambda}
\label{cinqueapp}
\eeqa
 
The sum in $V$ run over $I=1,\ldots,k$ distributed between the
Young tableaux $Y_\lambda$'s.
The first three terms between brackets in (\ref{trace}) 
come from the $U(k)$ adjoint fields $B_1,B_2,\chi_{\C}$ 
in (\ref{brsn2}). The $-1$ inside the bracket
comes from the Vandermonde determinant. The last two terms
are associated to the $I^\dagger$, $J^\dagger$ bifundamentals respectively.
After a long but straight algebra (see \cite{nakajima} for details) one finds
\cite{Flume:2002az}:    
\beqa
\chi
&= &\sum_{\lambda,\tilde{\lambda}}^N \sum_{s\in Y_j}
\left(T_{a_{\lambda\tilde{\lambda}}} T_1^{-h(s)} T_2^{v(s)+1}+
T_{ a_{\tilde{\lambda}\lambda}} T_1^{h(s)+1} T_2^{-v(s)}\right)
\label{tracef}
\eeqa
with
\beqa
h(s)=\nu_{i_\lambda}-j_\lambda \quad\quad v(s)=\tilde{\nu}^\prime_{j_\lambda}-i_\lambda
\eeqa
Notice that  $\tilde{\nu}^\prime_{j_\lambda}$ is defined only for $j_\lambda\leq\tilde{\nu}_{1_\lambda}$. 
For $j_\lambda > \tilde{\nu}_{1_\lambda}$ we take $\tilde{\nu}^\prime_{j_\lambda}=0$. 
 $h(s)$ ($v(s)$) is the number of black (white) circles in Fig.1.
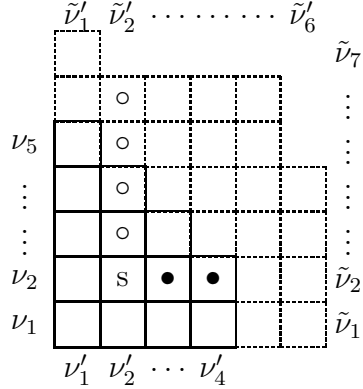
\begin{figure}[ht]
\label{figura}
\setlength{\unitlength}{2mm}
\begin{center}
\begin{picture}(25,25)(-5,-20)
\put(1.5,4){\makebox(0,0){$\tilde\nu^\prime_1$}}\put(4.5,4){\makebox(0,0){$\tilde\nu^\prime_2$}}
\put(7.5,4){\makebox(0,0){$\ldots$}}\put(10.5,4){\makebox(0,0){$\ldots$}}
\put(13.5,4){\makebox(0,0){$\ldots$}}\put(16.5,4){\makebox(0,0){$\tilde\nu^\prime_6$}}
\put(-2,-16.5){\makebox(0,0){$\nu_1$}}\put(-2,-13.5){\makebox(0,0){$\nu_2$}}
\put(-2,-10.5){\makebox(0,0){$\vdots$}}\put(-2,-7.5){\makebox(0,0){$\vdots$}}
\put(-2,-4.5){\makebox(0,0){$\nu_5$}}
\put(+1.5,-19.5){\makebox(0,0){$\nu^\prime_1$}}\put(+4.5,-19.5){\makebox(0,0){$\nu^\prime_2$}}
\put(+7.5,-19.5){\makebox(0,0){$\ldots$}}
\put(+10.5,-19.5){\makebox(0,0){$\nu^\prime_4$}}
\put(19.5,-16.5){\makebox(0,0){$\tilde\nu_1$}}\put(19.5,-13.5){\makebox(0,0){$\tilde\nu_2$}}
\put(19.5,-10.5){\makebox(0,0){$\vdots$}}\put(19.5,-7.5){\makebox(0,0){$\vdots$}}
\put(19.5,-4.5){\makebox(0,0){$\vdots$}}\put(19.5,-1.5){\makebox(0,0){$\vdots$}}\
\put(19.5,1.5){\makebox(0,0){$\tilde\nu_7$}}
\put(0,0){\dashbox{.2}(3,3)}
\put(0,-3){\dashbox{.2}(3,3)}\put(3,-3){\dashbox{.2}(3,3){$\circ$}}\put(6,-3){\dashbox{.2}(3,3)}
\put(9,-3){\dashbox{.2}(3,3)}\put(12,-3){\dashbox{.2}(3,3)}
\put(0,-6){\framebox (3,3)}\put(3,-6){\dashbox{.2}(3,3){$\circ$}}\put(6,-6){\dashbox{.2}(3,3)}
\put(9,-6){\dashbox{.2}(3,3)}\put(12,-6){\dashbox{.2}(3,3)}
\put(0,-9){\framebox (3,3)}\put(3,-9){\framebox (3,3){$\circ$}}\put(6,-9){\dashbox{.2}(3,3)}
\put(9,-9){\dashbox{.2}(3,3)}\put(12,-9){\dashbox{.2}(3,3)}\put(15,-9){\dashbox{.2}(3,3)}
\put(0,-12){\framebox (3,3)}\put(3,-12){\framebox (3,3){$\circ$}}\put(6,-12){\framebox (3,3)}
\put(9,-12){\dashbox{.2}(3,3)}\put(12,-12){\dashbox{.2}(3,3)}\put(15,-12){\dashbox{.2}(3,3)}
\put(0,-15){\framebox (3,3)}\put(3,-15){\framebox (3,3){s}}\put(6,-15){\framebox (3,3){$\bullet$}}
\put(9,-15){\framebox (3,3){$\bullet$}}\put(12,-15){\dashbox{.2}(3,3)}
\put(15,-15){\dashbox{.2}(3,3)}
\put(0,-18){\framebox (3,3)}\put(3,-18){\framebox (3,3)}\put(6,-18){\framebox (3,3)}
\put(9,-18){\framebox (3,3)}\put(12,-18){\dashbox{.2}(3,3)}\put(15,-18){\dashbox{.2}(3,3)}
\end{picture}
\caption{Two generic Young diagrams denoted by the indices $\lambda, \tilde\lambda$ in the main text.}
\end{center}
\end{figure}

The sum in (\ref{tracef}) runs over $2kN$ eigenvalues, the complex dimension of the moduli space. 
Moreover for generic $a_\lambda,\epsilon_1,\epsilon_2$ there are no zero eigenvalues
in (\ref{tracef}), since for $\lambda=\tilde{\lambda}$,
the quantities $h(s),v(s)$ are non-negative. Cancellations of zero
eigenvalues in (\ref{trace}) can be traced to the term 
$V^*\times V\times (T_1-1)(1-T_2)$. Zero eigenvalues are associated to the non-zero components of 
$B_1, B_2, I$ which we have discussed above.
Roughly the factor $(T_1-1)$ takes care of the cancellations connected to the $B_1$ components 
(horizontal neighbors)
and $(1-T_2)$ does the same for $B_2$
(vertical neighbors) in the sum in $V^*\times V\times (T_1-1)(1-T_2)$.
As anticipated, this mimics the reduction via ADHM constraints, since the negative 
contributions inside the brackets $-T_1T_2-1$ can be associated to
the fermionic ADHM constraints implemented by $\chi_{\C}, \chi_\R$ and to the $U(k)$ invariance.

Replacing the sum by a product over the eigenvalues we finally find:
\beqa
{\cal Z}_k=\sum_{x_0}  {1\over {\rm Sdet}
 \hat{{\cal L}}_{x_0}} &=&\sum_{\{Y_\lambda\}}
\prod_{\lambda,\tilde{\lambda}}^N \prod_{s\in Y_{\lambda}}{1\over E(s)(E(s)-\epsilon)}
\label{generalsdet}
\eeqa
with
\beqa
E(s) &=& a_{\lambda\tilde{\lambda}}-\epsilon_1 h(s)+\epsilon_2(v(s)+1)
\eeqa

\subsection{${\cal N}=2$ Supersymmetric Theories with fundamental matter}

In the presence of $N_F$ fundamental hypermultiplets,
the multi-instanton action gets a new contribution which can be
written as \cite{Fucito:2001ha}
\beq
S_{hyp} = - Q_{\epsilon} {\rm Tr} \left[ {h}^{\dagger}_f{\cal K}_f + {{\cal K}}^{\dagger}_f h_f \right] \ \ ,
\label{hyp}
\eeq
where $({\cal K}_f,h_f)$, $~f=1,..,N_F$ 
represent respectively 
the fermionic collective coordinates for the matter fields and
their bosonic auxiliary variables. 
They are all matrices transforming in the
$(\bar k, N_F)$ representation of the $U(k)\times U(N_F)$
group \footnote{We explicit the flavour index $f$ for convenience.}.
From the brane-engineering point of view, $({\cal K}_f,h_f)$ are the massless
excitation modes of open strings stretching between $k$ D(-1) and
$N_F$ D7 fractional branes.

The BRST transformations of these fields are given by 
\beqa
Q_{\epsilon} {\cal K}_f&=& h_f \quad\quad 
Q_{\epsilon}h_f = \phi{\cal K}_f + m_f {\cal K}_f \ \ , \nonumber\\
Q_{\epsilon} {\cal K}^{\dagger}_f&=& {h}^{\dagger}_f \quad\quad 
Q_{\epsilon}{h}^{\dagger}_f = -{\cal K}^{\dagger}_f \phi - m_f {\cal K}^{\dagger}_f \ \ ,
\label{Qhyp}
\eeqa
$m_f$ being the mass of the $f$-th flavour.
The vector field $Q^*$ generating the above $Q_{\epsilon}$ action in 
the supermoduli space is given by
\beq
Q^*=  h_f{\partial\over {\partial {\cal K}_f}} + (\phi + m_f ){\cal K}_f {\partial\over{\partial h_f}} \ \ ,
\label{qmf}
\eeq
which has to be added to the vector field (\ref{fermvecf}) for the pure ${\cal N}=2$ theory. 
The critical points are still given by (\ref{critical}), 
since the new components (\ref{qmf}) of the vector field are set to zero simply by imposing $h_f=0$.
From (\ref{qmf}) it follows that the contribution of each flavour $f$ to the supertrace is simply
\beq
\delta \chi= - T_{m_f}\times  V = 
- \sum_{\lambda}^N \sum_{s\in Y_{\lambda}} T_{a_{\lambda}} 
T_1^{-j_{\lambda}+1} T_2^{-i_{\lambda}+1} T_{m_f} \ \ ,
\label{strf}  
\eeq
with $T_{m_f}=e^{im_f}$ an element of the maximal torus $U(1)^{N_F}\subset U(N_F)$.
Taking into account the contribution of the $N_F$ hypermultiplets, (\ref{generalsdet})
becomes then 
\beq
{\cal Z}_k=\sum_{\{Y_\lambda\}}
\prod_{\lambda,\tilde{\lambda}}^N \prod_{s\in Y_{\lambda}}{F(s)\over E(s)(E(s)-\epsilon)} \ \ ,
\label{zfund}
\eeq
where we defined
\beq
F(s) = \prod_{f=1}^{N_F} (\varphi(s) + m_f) \ \ ,
\label{fs}
\eeq
with $\varphi(s)=\varphi_{i_\lambda j_\lambda}$ given by (\ref{solcritical}).

\subsection{${\cal N}=2^*$ Supersymmetric Theories with Gauge Group $SU(N)$}

Our techniques in the previous subsections can be
straightforwardly extended to the ${\cal N}=2^*$ case.
 As we mentioned before, one can identify the parameter $m$
in $U(1)_m$ with the mass of the ${\cal N}=2$
hypermultiplet. Notice that this identification was already 
implicit in our
${\cal N}=2$ analysis above since 
the fields projected out in the reduction ${\cal N}=4 \to {\cal N}=2$
are precisely those charged under $U(1)_m$.

The action of the BRST operatore, $Q^*$, is now given by:
\beqa\label{vecfield2*}
Q^* &&=\mu_I{\partial\over\partial I}+
\mu_J{\partial\over\partial J}+K{\partial\over\partial\mu_K}
+L{\partial\over\partial\mu_L}+
{\cal M}_\ell{\partial\over\partial B_\ell}
+H_v{\partial\over\partial \chi_v}+\eta{\partial\over\partial \bar\phi}
\nn\\
&&+(\phi -a)I{\partial\over\partial \mu_I}
+(-\phi +a+\epsilon )J{\partial\over\partial \mu_J}
+(\phi-a-m) \mu_K{\partial\over\partial K}
+(-\phi+a+\epsilon-m) \mu_L{\partial\over\partial L}\nn\\
&&+([\phi,B_\ell]+\lambda_\ell B_\ell)
{\partial\over\partial {\cal M}_\ell}
+([\phi,\chi_v]+\lambda_v\chi_v){\partial\over\partial H_v}+
[\phi,\bar\phi]{\partial\over\partial \eta}
\eeqa
The critical points are again given by (\ref{solcritical}) since for 
generic $m$ the condition $Q^*=0$ requires $B_3=B_4=H_v=0$.
Reading from $(\ref{vecfield2*})$ the spectrum of eigenvalues we find:
\beq \label{trace4}
\chi=(1-T_m^{-1})\left[
V^*\times V\times (T_1+T_2-T_1T_2-1)+ W^*\times V
+V^*\times W\times T_1T_2\right]
\eeq 
Remarkably the contributions of massive fields match that of 
the ${\cal N}=2$ in (\ref{trace}) but with eigenvalues shifted by $-m$.
The final result can then be written as\footnote{Once again, if we would plug 
(\ref{sdet}) in (\ref{generalsdetn=2*}) one would recover (3.25) 
in \cite{Nekrasov:2002qd}.}
\beq
{\cal Z}_k =\int \prod_{I=1}^k d\varphi_I{\prod_{I<J} \varphi^2_{IJ}\over {\rm Sdet}\,{\cal L}}=
\sum_{x_0}  {1\over {\rm Sdet}
 \hat{{\cal L}}_{x_0}} =\sum_{\{Y_\lambda\}}
\prod_{\lambda,\tilde{\lambda}=1}^N  
\prod_{s\in Y_\lambda}{(E(s)-m)(E(s)-\epsilon+m)\over E(s)(E(s)-\epsilon)}.
\label{generalsdetn=2*}
\eeq
Notice that now the superdeterminant reduce to a product over $2kN$ bosonic
and $2kN$ fermionic factors. Moreover, as in the ${\cal N}=2$ case, the
 superdeterminant is non-trivial due to the cancellation of zero
eigenvalues between bosons and fermions.

\subsubsection{Some explicit examples: k=1,2 }

Here, for the sake of completeness, we explicitly evaluate formula 
(\ref{generalsdetn=2*}) for k=1,2,
recovering the results in \cite{Nekrasov:2002qd,Flume:2002az}.
It is useful
to introduce the following definitions:
\beqa
f(x)&=&{(x-m)(x+m-\epsilon)\over x(x-\epsilon)}\quad\quad
g(x)={1\over x(x-\epsilon)}\nn\\
T_\alpha(x)&=&\prod_{\tilde{\alpha}\neq \alpha}\, f(a_{\alpha\tilde{\alpha}}+x)\quad\quad
S_\alpha(x)=\prod_{\tilde{\alpha}\neq \alpha} g(a_{\alpha\tilde{\alpha}}+x)
\label{fst}
\eeqa
In terms of these definitions we can rewrite:
\beq
{\cal Z}_k =\sum_{\{Y_\lambda\}}
\prod_{\lambda,\tilde{\lambda}=1}^N  
\prod_{s\in Y_\lambda}\, f(E(s))
\eeq

Let us start by considering the k=1 case:
$Y_\alpha= 
\raisebox{.1cm}{\framebox[.2cm]{}}$ , $Y_{\beta\neq \alpha}=\{ \emptyset \}$. From the above definitions 
we have $v(s)=h(s)=0$ for $\tilde{\lambda}=\alpha$ 
while $v(s)=-1,h(s)=0$ for $\tilde{\lambda}\neq \alpha$.
Summing up over diagrams of this kind one finds
\beq
Z_1=\sum_\alpha f(\epsilon_2) T_\alpha(0) 
\label{z1}
\eeq
For k=2 we have three diagrams: \\\\
I) $Y_\alpha=$ 
\raisebox{.1cm}{\framebox[.2cm]{}} , $Y_\beta=$ 
\raisebox{.1cm}{\framebox[.2cm]{}} , 
$Y_{\gamma\neq \alpha,\beta}=\{ \emptyset \}$:
\beq
Z_2^{I}={1\over 2}\sum_{\alpha\neq \beta}f(\epsilon_2)^2 f(a_{\alpha\beta}+\epsilon_2)f(a_{\beta\alpha}+\epsilon_2){T_\alpha(0)T_\beta(0)
\over f(a_{\alpha\beta}) f(a_{\beta\alpha})}
\eeq
The contribution ${T_\alpha(0)\over f(a_{\alpha\beta})}$ comes from the product in (\ref{generalsdetn=2*}) with
$\lambda=\alpha,\tilde{\lambda}\neq \alpha,\beta$ for which $h(s)=0,v(s)=-1$.
 The term 
$\lambda,\tilde{\lambda}=\alpha,\beta$, i.e.  $h(s)=v(s)=0$ 
gives $f(\epsilon_2)$ or $f(a_{\alpha\beta}+\epsilon_2)$
in the case of $\lambda=\tilde{\lambda}=\alpha$ and $\lambda=\alpha,\tilde{\lambda}=\beta$ respectively.
Similar contributions come from terms with $\alpha\leftrightarrow \beta$ exchanged. 

II) $Y_\alpha=$ 
\raisebox{.1cm}{\framebox[.2cm]{}\framebox[.2cm]{}}, $Y_{\tilde{\alpha}\neq \alpha}=\{ \emptyset \}$:
\beq
Z^{II}=\sum_\alpha  f(\epsilon_2) f(\epsilon_2-\epsilon_1) T_\alpha(0)T_\alpha(-\epsilon_1)
\label{z2}
\eeq
Now $ f(\epsilon_2) f(\epsilon_2-\epsilon_1)$ comes from the terms in
 (\ref{generalsdetn=2*}) with
$\lambda=\tilde{\lambda}=\alpha$ i.e. $v(s)=0,h(s)=0,1$, while  the product over $\tilde{\lambda}\neq \lambda=\alpha$,
 $v(s)=-1,h(s)=0,1$ brings the $T_\alpha$ contributions.
  
Finally the third diagram is the transposition of the one above and its contribution
can be read from (\ref{z2}) by exchanging $\epsilon_1\leftrightarrow \epsilon_2$. 
Setting $\epsilon_1=-\epsilon_2=\hbar$ and using the identification 
\cite{Moore:1997pc,Nekrasov:2002qd}:
\beq\label{genfunc}
Z(a,\epsilon_1,\epsilon_2)=\sum_k Z_kq^k=
\exp{({{\cal F}^{inst}\over\epsilon_1\epsilon_2})}.
\eeq
one recovers the
results in \cite{Chan:1999gj}:
\beqa
{\cal F}_1 &=&-\lim_{\hbar\to 0}
 \hbar^2 Z_1 = m^2\sum_\alpha T_\alpha\nn\\
{\cal F}_ 2&=&-\lim_{\hbar\to 0}
\hbar^2\left(Z_2-{1\over 2} Z_1^2 \right)=\sum_\alpha 
\left( {1\over 4} m^4 T_\alpha T^{''}_\alpha-{3\over 2} m^2 T_\alpha^2 \right)\nn\\
&&+m^4\,\sum_{\alpha\neq \beta}T_\alpha T_\beta
\left({1\over a_{\alpha\beta}^2}-{1\over 2(a_{\alpha\beta}-m)^2}-{1\over 2(a_{\alpha\beta}+m)^2}\right)
\eeqa
with  $T_\alpha=T_\alpha(0)$. 
The ${\cal N}=2$ analog of formulae (\ref{z1},\ref{z2})  can be simply obtained by replacing
$f(x),T_\alpha(x)$ by $g(x),S_\alpha(x)$ respectively given by (\ref{fst}).
 For $SU(2)$, formulae obtained in this way match those of \cite{Flume:2001kb}.

\subsection{${\cal N}=4$ Supersymmetric Theories with Gauge Group $SU(N)$}

This case can be easily deduced from (\ref{generalsdetn=2*}) by taking the limit $m\to 0$.
This limit gives $\hat{{\cal L}}_{x_0}=1$, thus applying (\ref{locth}) we get
\beq\label{finalen=4} 
{\cal Z}_k=\int_M e^{S^{N=4}}=\sum_{\{ k \}}\,1
\eeq
with $\{ k \}$ the partitions of $k$.
That is the partition function of ${\cal N}=4$ is the sum over all critical 
points of the vector field and it 
gives the Euler characteristic of the moduli space \cite{berline}. 
This correponds to the N-colored number of partitions of an integer $k$. 
Going now to the generating function we see that 
\beqa\label{sumdivi}
{\cal Z}=\sum_{k\ge 0}Z_k q^k&=& 
\sum_{k\ge 0}q^k \sum_{ \{ k_\lambda \}}1
=\prod_{n=1}^\infty{1\over (1-q^n)^N}
\eeqa
Defining 
\beq
{\cal F}=\sum_{k>0} q^k {\cal F}_k ={\rm ln}\, {\cal Z}=
N \sum_{n>0}\ln (1-q^n)=N\sum_{k>0}q^k\sum_{d|k}{1\over d}.
\eeq
one finds \footnote{A similar computation for the moduli space of instantons
of winding number $k=1/2$ on a Eguchi-Hanson manifold was carried out in \cite{Bianchi:1995ad}}
${\cal F}_k=N\sum_{d|k}1/d$. 
This result was already announced in \cite{Dorey:2000zq} and motivated in \cite{Dorey:2001ym}
on the basis of a reasoning coming from string theory: in \cite{Green:2000ke} the 
effective action of a single D3 brane  of the IIB theory at order $\alpha^{\prime 4}$ was computed. 
The coupling is  given by the modular invariant function $h(\tau,\bar\tau)=\ln|\tau_2\eta(\tau)^4|$.
By computing the generating functional of the instanton induced contributions to the scattering amplitude 
on the D3 brane and comparing with the results of \cite{Green:2000ke}, the
value of ${\cal F}_k$ is found. 
Here the result is recovered by a direct evaluation of the instanton 
contributions.

The fact that the result of (\ref{sumdivi}) is a function with particular properties
under modular transformations is a very satisfying feature. Multi-instanton calculus exactly reproduces
the important features of mathematical objects which have been studied with very different techniques.

\section*{Acknowledgements}
This work was supported in part by the EEC contracts HPRN-CT-2000-00122, HPRN-CT-2000-00131  and 
HPRN-CT-2000-00148, by the INTAS contract 99-0-590, by the MURST-COFIN
contracts 2001-025492 and 2000-02262971, and by a EEC Marie Curie Individual Fellowship
HPMF-CT-2001-01504.
The authors want to thank R.Flume and R.Poghossian for accepting an invitation to Rome and 
for many interesting discussions during their stay.
A.T. thanks N. Nekrasov for useful discussions.

\appendix
\section{Appendix} 
\setcounter{equation}{0}
Assume that $M$ is symplectic, 
with symplectic form $\omega$, that $G$ acts on $M$ by
symplectomorphisms (i.e., $g^\ast\omega=\omega$ for all $g\in G$), 
and that this action
admits a momentum map $\mu\colon M\to g^*$. 
Let 
$\alpha=\mu+\omega$. Here $deg(\alpha)=2$, since $\omega$ is a two-form and $\mu$ is a linear functional 
on $\g$,
see (\ref{deg}). Now,
\beq (g\cdot\alpha)(\xi)=g^*(\mu(ad_{\g^{-1}}\xi))+g^*\omega=\mu(\xi) +\omega=\alpha(\xi) \eeq  since $g$
acts as a symplectomorphism and $\mu$ is a momentum map. By definition, $\omega$ is closed and 
$\mu$ is a function, so that
\beq
(D\alpha)(\xi)=(d\alpha(\xi)-i_{\xi^*}\alpha(\xi))=(d(\mu+\omega)-i_{\xi^*}(\mu+\omega))=
d\mu(\xi) -i_{\xi^*}\omega=0 \ \ .
\label{dermommap}\eeq
It then follows that $\alpha=\mu+\omega$ is an equivariantly closed form
and the conditions of the localization formula are met.
Plugging in (\ref{locth}) we get the
Duistermaat-Heckman formula
\beq\label{dhth}
\int_X e^{\mu+\omega}=\int_X{\omega^{n/2}\over (n/2)!}e^\mu=(-2\pi)^{n/2}
\sum_{x_0}{\alpha_0(\xi)(x_0)\over 
{\rm det}^{1\over 2}\, {\cal L}_{x_0}}.
\eeq
since $\alpha_0(\xi)=e^\mu$.

Let now specify to our case where $G=U(k)\times U(1)^{N-1}\times 
T^2$. 
The condition of vanishing potential allows to 
take the Cartan part of the $U(N)$ algebra in $G$.
Given the symplectic form 
\beq\label{sympform}
\omega=dB_1\wedge dB_1^\dagger+dB_2\wedge dB_2^\dagger
+dI\wedge dI^\dagger-dJ^\dagger\wedge dJ=dx\wedge dx^\dagger
\eeq
and the component of the vector field 
\beq\label{compvec}
(Q^*)^i=(\phi I-I a,-J\phi +aJ+\epsilon J,[\phi,B_1]+\epsilon_1 B_1,[\phi,B_2]+\epsilon_2 B_2)
\eeq
we compute $d\mu=i_{Q^*}\omega=(Q^*)^i (dx^\dagger)^i$ from which
\beq
\mu=([\phi,B_1]+\epsilon_1B_1)B^\dagger_1+([\phi,B_2]+\epsilon_2B_2)B^\dagger_2+(\phi I-I a)I^\dagger+
J^\dagger(-J\phi +aJ+\epsilon J).
\eeq

How come that localization formulae with ``actions'' $\mu$ and $S^{{\cal N}=2}$ give the same results?
The answer lies in ${\cal L}_{x_0}$. To determine its eigenvalues the only information we need is to know the 
components of the vector field $Q^*$. 
There is no reference to any action. All the information is
encoded in the BRST transformations.

\section{Appendix}
\setcounter{equation}{0}
Here we collect some background material.
Following \cite{dhkmv} we decompose the quantum
numbers of the D(-1)-D3 system in terms of the 
reduced Euclidean Lorentz group $SO(6)\times SO(4)$. 
The ten dimensional spinor and gauge connection are taken as
\beqa\label{dimred}
\psi=\sqrt{\frac{\pi}{2}}\pmatrix{0\cr 1}\otimes
\pmatrix{{\cal M}^{ A}_\beta \cr 0}&+&
\sqrt{\frac{\pi}{2}}\pmatrix{1\cr 0}\otimes
\pmatrix{0 \cr\bar\lambda^A_{\dbeta}},\nonumber\\
\\
A_M&=&(\chi_a,a^\prime_n),\nonumber
\eeqa
while the low energy limit of the strings stretched between the $k$ Dp and $N$ D(p+4)-branes is given
by the fields $(w_{\dot\alpha},\mu^A;\bar w^{\dot\alpha},\bar\mu^A)$. We have denoted by $a=1,\ldots,6$
the indices of $SO(6)$, by $A=1,\ldots,4$ those of $SU(4)\cong SO(6)$ and by $\alpha, \dot\alpha=1,2$
those of $SO(4)\cong SU(2)\times SU(2)$.
The ADHM action of such system is given by \cite{dhkmv}
\beq
S_{k,N}={1\over g_0^{2}}S_{G} + S_{K}+S_{D}
\label{cometipare}
\eeq
with
\beqa\label{Sd}
&&S_{G}={\rm tr}_{k}\big(-[\chi_a,\chi_b]^2+\sqrt{2}i\pi
\lambda_{\dot{\alpha} A}[\chi_{AB}^\dagger,\bar\lambda^{\dot{\alpha}}_B]
-D^{c}D^{c}\big) \\
&&S_{K}=-{\rm tr}_{k}\big([\chi_a,a_{n}]^2
-\chi_a \bar{w}^{\dalpha}
w_{\dalpha}\chi_a + \sqrt{2}i\pi
{\cal M}^{\prime~\alpha A}[\chi_{AB}
{\cal M}^{\prime~B}_{\alpha}]-2\sqrt{2} i \pi
\chi_{AB} \bar{\mu}^{A} \mu^{B} \big)\, \nonumber \\
&&S_{D}={\rm tr}_k\,\big(i \pi\left(
-[a_{\alpha\dot{\alpha}},{\cal M}^{\prime~\alpha A}]
+\bar{\mu}^{A} w_{\dot{\alpha}}
+\bar{w}_{\dot{\alpha}}\mu^{A}\right)
\bar\lambda^{\dot{\alpha}}_{A}
+D^{c}\left(\bar{w} \tau^c w-i \bar{\eta}_{mn}^c [a_{m},a_{n}]
\right)\big)\nonumber
\eeqa
To go to an action with a lower number of 
supersymmetries it is sufficient to repeat the 
above construction in the case of fractional branes.
 A fractional brane lives at orbifold singularities and
its low energy field theory must be invariant under the action of the discrete group by which we mod the 
original space-time. 
The set of invariant fields is clearly smaller 
than the original one and the final
theory has thus less supersymmetries \cite{Fucito:2001ha}.
 To be consistent with the notation adopted
for the ADHM variables in the second chapter, we now set\footnote{We choose $\sigma^n_{\alpha\dot\alpha}=
(-1,i\tau^c)$.}
\beqa\label{conventions}
w_{\dot\alpha}&=&\pmatrix{I^\dagger\cr J},\nonumber\\
B_1&=&-a^\prime_0+ia^\prime_3 \quad\quad B_2=-a^\prime_2+ia^\prime_1,\label{convention}\\
B_3&=&{1\over \sqrt{2}}(-\chi_1+i\chi_4) \quad\quad B_4={1\over \sqrt{2}}(-\chi_2+i\chi_5).  
\label{phis}\\
\phi &=&{1\over \sqrt{2}} (-\chi_3+i \chi_6)\quad\quad\bar{\phi} ={1\over \sqrt{2}} (-\chi_3-i \chi_6) 
\label{phi}
\eeqa

Let's discuss the ${\cal N}=2^*$ case:
the fields in (\ref{phis}), together with the fermionic components given by $A=3, 4$ 
and some new auxiliary
fields $(K, L, H_2, H_3)$ give rise to the massive hypermultiplets with bosonic components
$(K, L, B_3, B_4, H_2, H_3)$ and fermionic components 
$(\mu^3, \mu^4, \bar\lambda^{\dot{\alpha}}_{3,4}, {\cal M}^{\prime~3,4}_{\alpha})~$.
By renaming $\mu^{3,4}\to\mu_{K,L}$, 
${\cal M}^{\prime~3,4}_{\alpha}\to \chi_{2,3}$
and $ \bar\lambda^{\dot{\alpha}}_{3,4}\to {\cal M}_{3,4}$
we finally get
the fields entering the action (\ref{N=2*act}) and transforming as (\ref{brsn4})
with $a=\epsilon=0$. Notice that upon the rescalings
$(I,J^{\dagger}, B_1, B_2)\rightarrow g_0^{1/2}(I,J^{\dagger}, B_1, B_2)$ 
and
$(B_3, B_4, \phi)\rightarrow g_0^{-1/2}(B_3, B_4, \phi)$ and integration on the auxiliary
fields $(\vec \chi, \vec H)$, the action (\ref{N=2*act}) reproduces (\ref{cometipare}),
integrated with respect to $(\bar\lambda^{\dot \alpha}_A, D^c)$.

We now specialize the above discussion to the case of fractional branes.
After the $\Z_2$ projection the multi-instanton
action can be read from (\ref{Sd}) with fermionic
indices $A,B$ now restricted to $\dot{A},\dot{B}=1,2$ 
(in the fundamental of the automorphism group) which corresponds to set the entire massive hypermultiplet to zero see (\ref{riduz}).
The action thus obtained can be seen as the implementation \`a la BRST of the ADHM constraints,
which we ''twist'' by identifying $\dot{A}$ with $\dot\alpha$.
The constraints are now given by (\ref{mommap2}).
To them we associate the doublet of auxiliary fields $(\bar\lambda^A_{\dot\alpha},D^c)$ in 
(\ref{Sd}) which we rename $(\chi=\chi_\R,\chi_\C; H=H_\R, H_\C)$ and the doublet
$(\bar{\phi}, \eta)$. Given all this, after introducing a v.e.v. for the scalar field we get
the action of ${\cal N}=2$ SYM \cite{Fucito:2001ha} 
\beqa\label{actionn=2}
S^{N=2}&=&QTr\{[\mu_I(I^\dagger\bar\phi-\bar aI^\dagger)+\mu_J(\bar\phi J^\dagger-J^\dagger\bar a)
+{\cal M}_{\hat\ell}[\bar\phi,B^\dagger_{\hat\ell}]]+h.c.\nonumber\\
&+&\chi_\R{\cal E}_\R+\chi_\C{\cal E}_\C+{1\over g_0^2}
(\eta[\phi,\bar\phi]+\chi\cdot H)\},
\eeqa
invariant under the BRST transformations (\ref{brsn2}) with $\epsilon_\ell=0$.

We are now ready to discuss the properties of (\ref{actionn=2}). For the sake of clarity let us, 
for the moment, disregard the auxiliary fields implementing the constraints by setting $H_\R, H_\C$
and their fermionic partners to zero. Moreover notice that the transformations (\ref{brsn2}) 
have been improperly
called BRST, since they do not square to zero. 
The complete BRST operator on the ADHM space
has been constructed in
\cite{Bellisai:2000bc} and reads
\beqa\label{brs1}
sI&=&\mu_I-CI\ \;\qquad s\mu_I=\phi I-I a -C\mu_I \ \,\nonumber\\
sJ&=&\mu_J+JC\ \;\qquad s\mu_J=-J\phi +aJ +\mu_J C ,\nonumber\\
sB_1&=&{\cal M}_1-[C,B_1] \ \;\qquad s{\cal M}_1=[\phi,B_1]-[C,{\cal M}_1]\ \,\nonumber\\
sB_2&=&{\cal M}_2-[C,B_2] \ \;\qquad s{\cal M}_2=[\phi,B_2]-[C,{\cal M}_2]\ \ ,\nonumber\\
s\phi&=&-[C\phi]\ \;\qquad sC=(\phi-a)-[C,C]\ \,
\eeqa
where $C$ is a $U(k)$ connection acting on the fields as 
\beq\label{nonso}
C\cdot(I,J,B_1,B_2)=(CI,-JC,[C,B_1],[C,B_2]).
\eeq
The $Q$ operator correspond to the covariant derivative on the ADHM moduli space $Q=s+C\cdot\,$.
In terms of the BRST operator $s$, the action (\ref{actionn=2}) can be written as 
\beq\label{act1}
S^{N=2}= s~Tr(\mu_I\bar aI^\dagger+J^\dagger\bar a\mu_J+ h.c.)= s~\alpha=Q\alpha \ \ .
\eeq
According to the grading (\ref{deg}), $\alpha$ is a 3-form and 
since it is $U(k)\times U(N)$ equivariant (see (4.39) in \cite{Bellisai:2000bc})
the actions of $s$ and $Q$ on it give the same result.
By substituting the fermions with their expressions (\ref{brs1}), $\alpha$ becomes a bosonic form
on the moduli space of instantons as suggested in \cite{Bellisai:2000bc} and since $s^2=0$, 
the BRST operator can be interpreted as a {\it bona fide} derivative. In \cite{Flume:2001kb} it is
suggested that, since the action of $s$ and Q on $\alpha$ are the same, it can be more convenient 
to interpret Q as the equivariant derivative D introduced earlier in (\ref{der}) and drop the connection $C$.
This does not mean to drop $C$ altogether but only when it acts on the form $\alpha$.
In \cite{Bellisai:2000bc} it is, in fact, shown that the presence of $C$ is crucial to recover the correct 
measure on the instanton moduli space. 
 
It is immediate to see that, due to BRST invariance, $\alpha$ is equivariantly closed and that the 
infinitesimal action of the bosonic vector field $\xi^*$ can be read from the action of $Q^2$ (\ref{der2}) 
which is the Lie derivative.
The localization theorems could now be applied. The bosonic part of the action (the part of the action
which is a zero form with respect to differentials of the ADHM variables) is given by $i_{\xi^*}\alpha$
which is  positive semi-definite. Then its zeroes are the critical points which could also be obtained
by computing the $Q^2$ on the bosonic variables 
\beqa\label{critpoint1}
Q^2I&=&\phi I-I a\ \,\nonumber\\
Q^2J&=&-J\phi +aJ \ \,\nonumber\\
Q^2B_l&=&[\phi,B_l]\ \ .
\eeqa
These are exactly the critical points found in \cite{Hollowood:2002ds}. As in that paper these critical
points are rather critical surfaces and the application of the localization theorem is rather cumbersome.
The useful suggestion now comes from \cite{Moore:1997dj,Moore:1998et,Nekrasov:2002qd}: we can introduce
a further symmetry in the problem which, without changing the cohomology, can ``reduce'' the critical 
surfaces to isolated critical points. It acts on the coordinate of spacetime as two independent rotations
in the $x_1,x_2$ and $x_3,x_4$ planes. The group element describing such rotations is $T^2=(t_1,t_2)$
with $t_i=\exp{i\epsilon_i},\, i=1,2$ acting on the complex coordinates as $(z_1,z_2)\to(t_1z_1,t_2z_2)$.
There is clearly a wide margin of arbitrarity in the choice of the parametrization of the $T^2$, since 
we can always arbitrarily rescale the complex coordinates. The only condition is to leave (\ref{mommap2}) invariant. Our action and BRST transformation must 
be changed accordingly to accomodate for the new symmetry. The resulting 
action and BRST transformations are described in the main body of the paper.

\section{Appendix}
\setcounter{equation}{0}
In this section we will comment on Proposition 5.8 in \cite{nakajima} which gives the character, 
$T_Z(\C^2)^{[n]}$, at the fixed points, $Z$, of the tangent space to the Hilbert scheme $(\C^2)^{[n]}$ 
of $n$ points on $\C^2$. 
With a little generalization, from this formula one 
can extract the eigenvalues of $Sdet$
in (\ref{generalsdet}), (\ref{generalsdetn=2*}). Here we will try to ``translate'' the setting 
of \cite{nakajima} in the language we have used for this paper.

From the definition of the map ${\cal L}_{x_0}$ 
it should become clear why we are after the   
character (or eigenvalues) of the tangent space. 

Now consider the problem, given a self-dual field strength $F_{\mu\nu}$ and a vector potential $A_\mu$
of investigating the infinitesimal variations $\delta A_\mu$ preserving the self-duality of 
$F_{\mu\nu}$. Then 
\beq
(d_2\delta A)_{\mu\nu}=\Pi_{\mu\nu}{}^{\alpha\beta}(D_\alpha\delta A_\beta-D_\beta \delta A_\alpha)=0.
\label{zeroapp}
\eeq
$D_\alpha$ is the gauge covariant derivative and $\Pi_{\mu\nu}{}^{\alpha\beta}$ the projector on the 
anti-self-dual part of a tensor. Among the solutions of (\ref{zeroapp}), there are those arising  from 
infinitesimal gauge transformations, $\varepsilon$, of the gauge field. They are of the form
\beq
(d_1\varepsilon)_{\mu}=D_\mu\varepsilon.
\label{zerozeroapp}
\eeq
Two solutions of (\ref{zeroapp}) are gauge equivalent if they differ by a field of the form  
(\ref{zerozeroapp}). The problem of finding the number of gauge inequivalent solutions to (\ref{zeroapp})
is more conveniently treated by representing the tangent space $T_Z(\C^2)^{[n]}$
as the quotient Ker $d_2$/Im $d_1$ associated to the complex
\beq\label{dueapp}
Hom(V,V)\quad{\buildrel d_1\over\longrightarrow} \matrix{Hom(V,Q\otimes V)\cr\oplus\cr Hom(W,V)\cr\oplus\cr 
Hom(V,\bigwedge^2Q\otimes W)}
{\buildrel d_2\over\longrightarrow}\quad Hom(V,V)\otimes\bigwedge{}^2Q
\eeq
introduced  to prove Proposition 5.8 in \cite{nakajima}. 
In (\ref{dueapp}) the symmetry with respect to the action of the two torus 
$T^2=(t_1,t_2)$ has been taken into account by introducing the doublet $Q$. 
$\bigwedge{}^2Q=t_1t_2=\exp{\{i\epsilon\}}$ is the totally antisymmetric combination which that is 
the determinant. The correspondence between this notation and that of the rest of the paper is
\beqa\label{treapp}
B_l\ \ &:& \ \  Hom(V,Q\otimes V)\nonumber\\
I^\dagger\ \ &:& \ \ Hom(W,V)\nonumber \\
J^\dagger \ \ &:& \ \ Hom(V,\bigwedge{}^2Q\otimes W)\nonumber \\
\chi_\R \ \ &:& \ \ Hom(V,V)\nonumber \\
\chi_\C \ \ &:& \ \ Hom(V,V)\otimes\bigwedge{}^2Q.\nonumber \\
\eeqa
Then
\beqa\label{quattroapp}
T_Z &=&Hom(V,Q\otimes V)+Hom(W,V)+Hom(V,\bigwedge{}^2Q\otimes W)-\nonumber\\
&&Hom(V,V)-
Hom(V,V)\otimes\bigwedge{}^2Q=V^*\otimes V\otimes(Q-\bigwedge{}^2Q\otimes W-1)\nonumber\\
&&+W^*\otimes V+V^*\otimes W\otimes\bigwedge{}^2Q.
\eeqa
At the critical point the tangent space can be decomposed in terms of the quantum numbers of 
$T^2\times U(1)^{n-1}$ giving (\ref{Vs}).

\end{document}